\pdfoutput=1
\documentclass[reprint,prl,twocolumn,showpacs,epsfig,longeq,superscriptaddress]{revtex4}
\usepackage{dcolumn}
\usepackage{bm}
\usepackage{amsbsy}
\usepackage{graphicx}
\usepackage{esint}
\providecommand{\tabularnewline}{\\}
\newcommand{\apjl}{ApJL}

\newcommand{\aap}{A~\&~A}
\newcommand{\mnras}{MNRAS}
\newcommand{\solphys}{Sol.Phys.}
\newcommand{\physrep}{Phys. Rep.}

\usepackage[english]{babel}

\begin{document}

\title{Helicity conservation in nonlinear mean-field solar dynamo}

\author{V.V. Pipin$^{1,2}$, D.D. Sokoloff$^{1,3}$,H.Zhang$^{1}$,  K.M. Kuzanyan$^{1,4}$}

\affiliation{ $^{1}$ Key Laboratory of Solar Activity, National Astronomical Observatories,
Chinese Academy of Sciences, Beijing 100012, China \\
$^{2}$Institute of Solar-Terrestrial Physics, Russian Academy of
Sciences, Irkutsk, 664033, Russia \\
 $^{3}$ Department of Physics, Moscow University, 119992 Moscow,
Russia\\
 $^{4}$ IZMIRAN, Russian Academy of Sciences, Troitsk, Moscow region
142190, Russia }

%\pacs{}

%\draft

\begin{abstract}
We explore the impact of magnetic helicity conservation on the mean-field
solar dynamo using the axisymmetric dynamo model which includes the
subsurface shear. Our results support the recent findings by \citet{2012ApJ...748...51H},
who suggested that the catastrophic quenching in the mean-field dynamo
is alleviated if conservation of the total magnetic helicity is taken
into account. We show that the solar dynamo can operate in the wide
rage of the magnetic Reynolds number up to $10^{6}$. We also found
that the boundary conditions for the magnetic helicity influence the
distribution of the $\alpha$-effect near the solar surface.

\end{abstract}
\maketitle

\section{Introduction}

The basic idea for the solar dynamo action was developed by \citet{par55}.
He suggested that the toroidal component of the magnetic field of
the Sun is stretched from the poloidal component by the differential
rotation ($\Omega$ effect) and the cyclonic motions ($\alpha$ effect)
return the part of the toroidal magnetic field energy back to the
poloidal component. This is the so-called $\alpha\Omega$ scenario.
This mechanism is implemented in the wide range of the solar dynamo
models (see review by \citealp{chrev05}).

The effect of the turbulence in the mean-field dynamo is represented
by the mean electromotive force $\boldsymbol{\mathcal{E}}\mathbf{=}\overline{\mathbf{u}\times\mathbf{b}}$,
where $\mathbf{u}$ and $\mathbf{b}$ are the fluctuating velocity
and magnetic fields. In the simplest case it can be found that $\boldsymbol{\mathcal{E}}=\alpha_{0}\overline{\mathbf{B}}+\left(\mathbf{\overline{V}}^{(p)}\times\overline{\mathbf{B}}\right)-\eta_{T}\left(\nabla\times\overline{\mathbf{B}}\right),$
where $\alpha_{0}$ is the $\alpha$ effect, $\mathbf{\overline{V}}^{(p)}$
is the turbulent pumping and $\eta_{T}$ is the turbulent diffusivity
\citep{krarad80}. The $\alpha$ effect is a pseudo-scalar (lacks
the mirror symmetry) which is related to the kinetic helicity of the
small-scale flows, i.e., ${\displaystyle \alpha_{0}=-\frac{\tau_{c}}{3}\overline{\mathbf{u}\cdot\boldsymbol{\nabla\times}\mathbf{u}}}$,
where $\tau_{c}$ is correlation time of turbulent motion. \citet{pouquet-al:1975b}
showed that the $\alpha$ effect is produced not only by kinetic helicity
but also by current helicity, and it is ${\displaystyle \alpha_{0}=-\frac{\tau_{c}}{3}\left(\overline{\mathbf{u}\cdot\boldsymbol{\nabla\times}\mathbf{u}}-\frac{\mathbf{\overline{b\cdot\boldsymbol{\nabla\times b}}}}{2\mu\overline{\rho}}\right)}$.
The latter effect can be interpreted as resistance of magnetic fields
against to the twist by helical motions. It leads to the concept of
the catastrophic quenching of the $\alpha$ effect by the generated
large-scale magnetic field. It was found that ${\displaystyle \alpha_{0}\left(\overline{B}\right)=\frac{\alpha_{0}\left(0\right)}{1+R_{m}\left(\overline{B}/\overline{B}_{eq}\right)^{2}}}$,
where $R_{m}$ is magnetic Reynolds number (see, \citealp{kle-rog99}
and references therein). In case of $R_{m}\gg1$, the $\alpha$ effect
is quickly saturated for the large-scale magnetic field strength that
is much below the equipartition value ${\displaystyle \overline{B}_{eq}\sim\sqrt{\overline{\rho}\mu_{0}\overline{u^{2}}}}$.
The result was confirmed by the direct numerical simulations (DNS)
\citep{oss2001}. The catastrophic quenching (CQ) is related to the
dynamical quenching of the $\alpha$ effect. It is based on conservation
of the magnetic helicity, $\overline{\chi}=\overline{\mathbf{a\cdot}\mathbf{b}}$
($\mathbf{a}$ is fluctuating part of the vector potential) and the
relation between the current and magnetic helicities $h_{\mathcal{C}}=\mathbf{\overline{b\cdot\boldsymbol{\nabla\times b}}\sim\overline{\chi}}/\ell^{2}$,
which is valid for the isotropic turbulence\citep{moff:78}. The evolution
equation for $\overline{\chi}$ can be obtained from equations that
govern $\mathbf{a}$ and $\mathbf{b}$, it reads as follows \citep{kle-rog99}:
\begin{eqnarray}
\frac{\partial\overline{\chi}}{\partial t} & = & -2\left(\boldsymbol{\mathcal{E}}\cdot\overline{\bm{B}}\right)-\frac{\overline{\chi}}{R_{m}\tau_{c}}-\boldsymbol{\nabla}\cdot\boldsymbol{\boldsymbol{\mathcal{F}}}^{\chi}-\eta\overline{\mathbf{B}}\cdot\mathbf{\overline{J}},\label{eq:hel}
\end{eqnarray}
where, in following to \citet{kle-rog99}, we introduce the helicity
fluxes $\boldsymbol{\boldsymbol{\mathcal{F}}}^{\chi}=\mathbf{\overline{a\times u}}\times\mathbf{B}-\mathbf{\overline{a\times(u\times b)}}$.
The helicity fluxes are capable to alleviate the catastrophic quenching
(see, e.g., \citealp{2005PhR...417....1B,vish-ch:01} and references therein).
Generally, it was found that the diffusive fluxes, which are $\sim\eta_{\chi}\boldsymbol{\nabla}\overline{\chi}$,
where $\eta_{\chi}$ is the turbulent diffusivity of the magnetic
helicity, work robustly in the mean-field dynamo models but it requires
$\eta_{\chi}>\eta_{T}$ to reach $\left|\overline{B}\right|\ge0.1\overline{B}_{eq}$.

Another possibility to alleviate the catastrophic quenching is related
with the non-local formulation of the mean-electromotive force \citep{2008A&A...482..739B}.
In fact, the Babcock-Leighton (BL) type dynamo is the the special
case of the mean-field dynamo with the nonlocal $\alpha$ effect.
\citet{2011AstL...37..286K} found that the nonlocal $\alpha$ effect
and the diamagnetic pumping can alleviate the catastrophic quenching.
The results by \citet{2007NJPh....9..305B} show that the strength
of the quenching can depend on the model design. Therefore, the problem
of the catastrophic quenching is actual for different types of the
solar dynamo.

Recently, \citet{2012ApJ...748...51H} revisited the concept of catastrophic
quenching and showed that for the shearing dynamos the Eq.(\ref{eq:hel})
produces the nonphysical fluxes of the magnetic helicity over the
spatial scales. They suggested to cure the problem using the global
conservation law for the total magnetic helicity that can be written
as follows: 
\begin{eqnarray}
\frac{d}{dt}\int\left\{
  \overline{\chi}+\overline{\mathbf{A}}\cdot\overline{\mathbf{B}}\right\}
dV&=&-\eta\int\left\{
  \overline{\mathbf{B}}\cdot\mathbf{\overline{J}}+\overline{\mathbf{b\cdot
      j}}\right\} dV\label{eq:int-cons}\\
&&-\int\boldsymbol{\nabla\cdot}\boldsymbol{\boldsymbol{\mathcal{F}}}^{\chi}dV\nonumber
\end{eqnarray}
where integration is done over the volume that comprises the ensemble
of the small-scale fields. We assume that $\boldsymbol{\boldsymbol{\mathcal{F}}}^{\chi}$
is the diffusive flux of the total helicity which is resulted from
the turbulent motions. Ignoring the effect of the meridional circulation
we write the local version of the Eq.(\ref{eq:int-cons}) as follows
\citep{2012ApJ...748...51H}:

\begin{equation}
\frac{\partial\overline{\chi}}{\partial t}=-\frac{\partial\left(\overline{\mathbf{A}}\cdot\overline{\mathbf{B}}\right)}{\partial t}-\frac{\overline{\chi}}{R_{m}\tau_{c}}-\eta\overline{\mathbf{B}}\cdot\mathbf{\overline{J}}-\boldsymbol{\nabla\cdot}\boldsymbol{\boldsymbol{\mathcal{F}}}^{\chi},\label{eq:helcon2}
\end{equation}
where ${\boldsymbol{\mathcal{F}}}^{\chi}=
-\eta_{\chi}\boldsymbol{\nabla}\left(\overline{\chi}+\overline{\mathbf{A}}\cdot\overline{\mathbf{B}}\right)$.
In the paper we employ this equation for the dynamical quenching of
the $\alpha$ effect in the solar dynamo model and show how it works
in the range of the magnetic Reynolds number $R_{m}=10^{3-6}$ those
are typical for the astrophysical conditions.

\section{Basic equations}

We study the mean-field induction equation in a perfectly conducting
medium: 
\begin{equation}
\frac{\partial\overline{\bm{B}}}{\partial
  t}=\boldsymbol{\nabla}\times\left(\boldsymbol{\mathcal{E}}
+\overline{\bm{U}}\times\overline{\bm{B}}\right),\label{eq:dyn}
\end{equation}
where $\boldsymbol{\mathcal{E}}=\overline{\bm{u\times b}}$ is the
mean electromotive force, with $\bm{u,\, b}$ being fluctuating velocity
and magnetic field, respectively, $\overline{\bm{U}}$ is the mean
velocity (differential rotation), and the axisymmetric magnetic field
is: 
\[
\overline{\bm{B}}=\bm{e}_{\phi}B+\nabla\times\frac{A\bm{e}_{\phi}}{r\sin\theta},
\]
where $\theta$ is the polar angle. The mean electromotive force $\boldsymbol{\mathcal{E}}$
is given by \citet{pi08Gafd}. It is expressed as follows: 
\begin{equation}
\mathcal{E}_{i}=\left(\alpha_{ij}+\gamma_{ij}^{(\Lambda)}\right)\overline{B}_{j}
-\eta_{ijk}\nabla_{j}\overline{B}_{k}.\label{eq:EMF-1}
\end{equation}
The tensor $\alpha_{ij}$ represents the $\alpha$-effect, $\gamma_{ij}^{(\Lambda)}$
is the turbulent pumping, and $\eta_{ijk}$ is the diffusivity tensor.
The $\alpha$ effect includes hydrodynamic and magnetic helicity contributions,
\begin{eqnarray}
\alpha_{ij} & = & C_{\alpha}\sin^{2}\theta\alpha_{ij}^{(H)}+\alpha_{ij}^{(M)}\label{alp2d}
\end{eqnarray}
The details in expressions for the kinetic part of the $\alpha$ effect
$\alpha_{ij}^{(H)}$, as well as $\gamma_{ij}^{(\Lambda)}$ and $\eta_{ijk}$
can be found in \citep{pipea2012AA}. The contribution of small-scale
magnetic helicity $\overline{\chi}=\overline{\bm{a\cdot}\bm{b}}$
($\bm{a}$ is the fluctuating vector-potential of the magnetic field)
to the $\alpha$-effect is defined as 
\begin{equation}
\alpha_{ij}^{(M)}=2f_{2}^{(a)}\delta_{ij}\frac{\overline{\chi}\tau_{c}}{\mu_{0}\overline{\rho}\ell^{2}}
-2f_{1}^{(a)}e_{i}e_{j}\frac{\overline{\chi}\tau_{c}}{\mu_{0}\overline{\rho}\ell^{2}}.\label{alpM}
\end{equation}
The nonlinear feedback of the large-scale magnetic field to the $\alpha$-effect
is described by a dynamical quenching due to the constraint of magnetic
helicity conservation given by Eq.(\ref{eq:helcon2}). The effect
of turbulent diffusivity, which is anisotropic due to the Coriolis
force, is given by: 
\begin{eqnarray}
\eta_{ijk}&=&3\eta_{T}\left\{ \left(2f_{1}^{(a)}
-f_{2}^{(d)}\right)\varepsilon_{ijk}-2f_{1}^{(a)}e_{i}e_{n}\varepsilon_{njk}\right.\label{eq:diff}\\
&&\left. +C_{\delta}f_{4}^{(d)}e_{j}\delta_{ik}\right\}, \nonumber
\end{eqnarray}
where functions $f_{\{1,2,4\}}^{(a,d)}$ depend on the Coriolis number.
They can be found in \citet{pi08Gafd}. The last part of Eq.~(\ref{eq:diff})
stands for the $\Omega\times J$ effect \citep{rad69}. The DNS dynamo
experiments support the existence of the dynamo effects induced by
the large-scale current and global rotation \citep{2008A&A...491..353K,2011A&A...533A.108S}.

We matched the potential field outside and the perfect conductivity
at the bottom boundary with the standard boundary conditions. For
the magnetic helicity we employ $\bar{\chi}=0$ at the bottom of the
convection zone. At the top we use two types of the boundary conditions
like 
\begin{eqnarray}
\nabla_{r}\bar{\chi} & = & 0,\label{eq:bc1}\\
\nabla_{r}\left(\bar{\chi}+\overline{\mathbf{A}}\cdot\overline{\mathbf{B}}\right) & = & 0.\label{eq:bc2}
\end{eqnarray}
To evolve the Eq.(\ref{eq:helcon2}) we have to define the large-scale
vector potential for each time-step. For the axisymmetric large-scale
magnetic fields where the vector-potential is 
$\overline{\mathbf{A}}=
\mathbf{e}_{\phi} A/ \left(r\sin\theta\right)+r\mathbf{e}_{r}T$.
% \begin{equation}
% \overline{\mathbf{A}}=
% \mathbf{e}_{\phi}P+\mathbf{r}T=\frac{\mathbf{e}_{\phi}}{r\sin\theta}A+r\mathbf{e}_{r}T.\label{eq:potent-1}
% \end{equation}
The toroidal part of the vector potential is governed by the dynamo
equations. The poloidal part of the vector potential can be restored
from equation
$\boldsymbol{\nabla}\times\left(\mathbf{r}T\right)=\mathbf{e}_{\phi}B$.
The restoration procedure is trivial for the pseudo-spectral numerical
schemes which are based on the Legendre polynomial decomposition for
the latitude profile of the large-scale toroidal field.
\begin{figure}
\begin{centering}
\includegraphics[width=0.99\columnwidth]{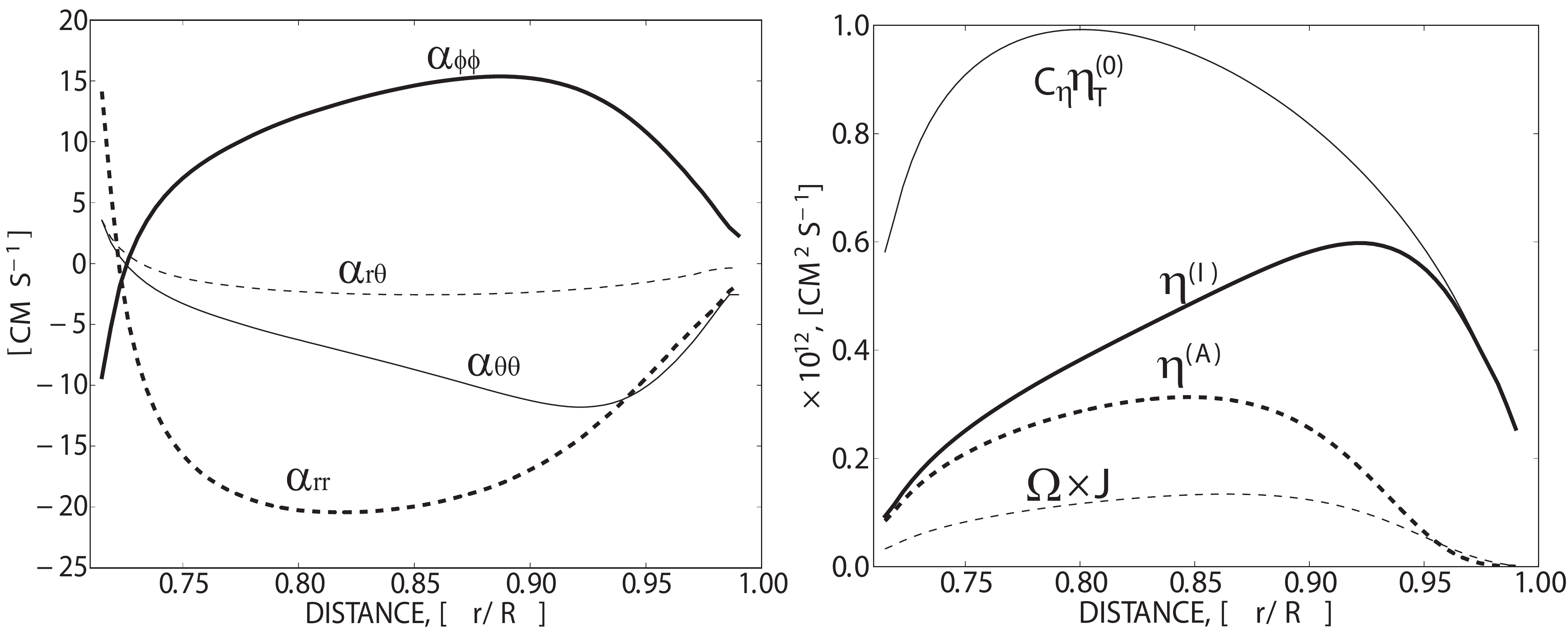}
\par\end{centering}
\caption{Left, (a), the profiles of the $\alpha$ effects components for the $\theta=45^{\circ}$.
Right, (b), the profiles of the background turbulent diffusivity $C_{\eta}\eta_{T}$,
the isotropic, $\eta^{(I)}$, and anisotropic, $\eta^{(A)}$, parts
of the magnetic diffusivity and $\Omega\times J$ effect (also known
as $\delta$ effect \citep{rad69}\label{fig:profiles}.}
\end{figure}
The choice of parameters in the dynamo is justified by our previous
studies \citep{pk11mf}. Here we use
$C_{\alpha}=0.03$,  $C_{\delta}=\frac{1}{3}C_{\alpha}$ and the diffusivity dilution factor $C_{\eta}=0.1$.
 The parameters of the models are summarized in the Table \ref{tab:Param}.
% , where it was shown that
% solar-types dynamos can be obtained for $C_{\alpha}/C_{\delta}>2$.
% In those papers we find the approximate threshold to be
%for a given diffusivity dilution factor of $C_{\eta}=0.1$.
% $C_{\alpha}\approx0.03$,  $C_{\delta}=\frac{1}{3}C_{\alpha}$ and the diffusivity dilution factor $C_{\eta}=0.1$.
%  The parameters
% of the models are summarized in the Table \ref{tab:Param}. 
\begin{table}
\caption{Summary of parameters of the models. Here, QT stands for the
  quenching type and  BC -  the boundary conditions.\label{tab:Param}}
\begin{centering}
\begin{tabular}{cccc}
\hline 
%|c|c|c|c|}
Model  & QT1$_{1}$, QT2$_{1}$  & QT1$_{2,3}$  & QT2$_{2,3}$\tabularnewline
\hline 
\hline 
$R_{\chi}$  & $10^{6}$  & $10^{3}$,$10^{5}$  & $10^{3}$,$10^{5}$\tabularnewline
\hline 
QT  & Eq.~(\ref{eq:hel-1}) ,Eq.~(\ref{eq:helcon2})  & Eq.~(\ref{eq:helcon2})  & Eq.~(\ref{eq:hel})\tabularnewline
\hline 
$\eta_{\chi}$  & $10^{-5}\eta_{T}$  & $10^{-2}\eta_{T}$  &  $10^{-2}\eta_{T}$  \tabularnewline
\hline 
BC  & Eq.~(\ref{eq:bc1})  & Eq.~(\ref{eq:bc1})  & Eq.~(\ref{eq:bc1}),Eq.~(\ref{eq:bc2})\tabularnewline
\hline 
\end{tabular}
\par\end{centering}
\end{table}

We estimate the turbulent parameters in the model on the base of the
mixing-length theory and use as the reference the solar interior model
computed by \citet{stix:02}. The differential rotation profile is
like that suggested by \citet{pk11apjl}. Fig.~\ref{fig:profiles}
shows the radial profiles of the $\alpha$ effect components and profiles
of the background turbulent diffusivity $C_{\eta}\eta_{T}$, the isotropic,
$\eta^{(I)}$, and anisotropic, $\eta^{(A)}$, parts of the magnetic
diffusivity as well as profile for the $\Omega\times J$ effect.
\begin{figure}[tv]
\includegraphics[width=0.99\columnwidth]{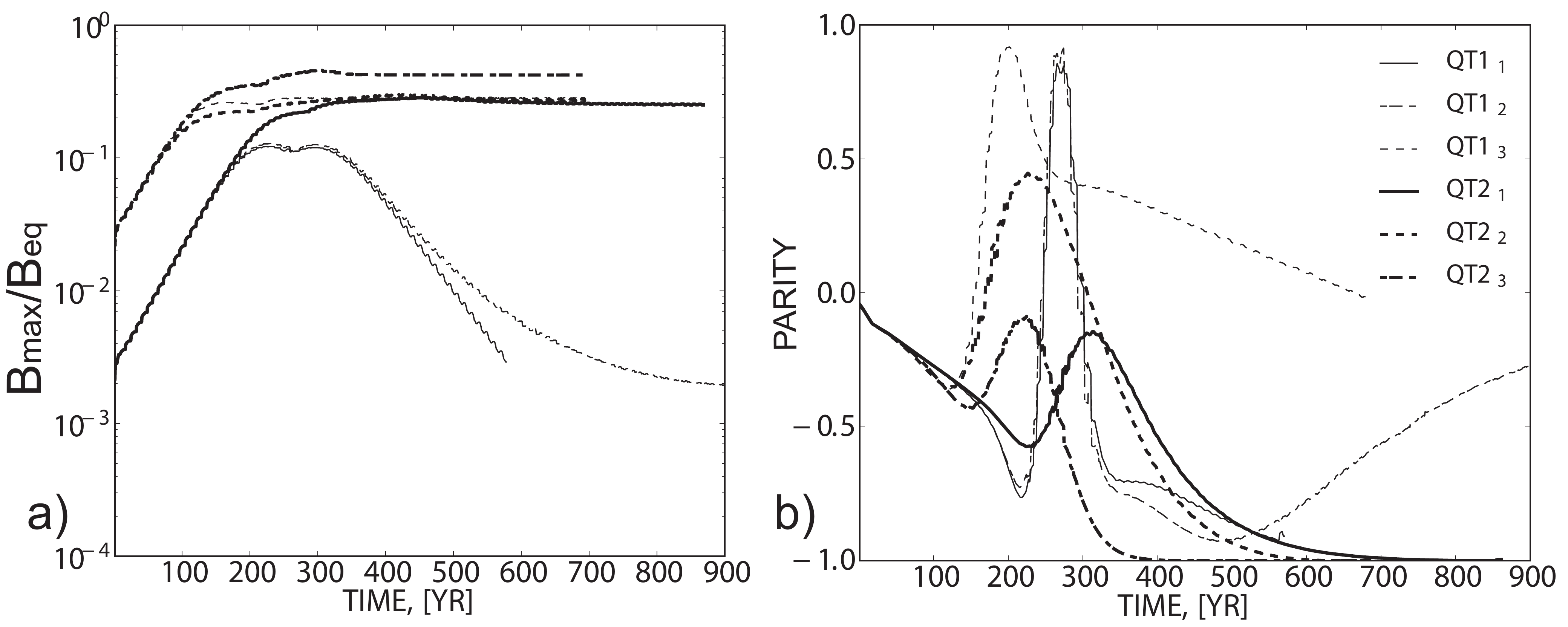}
\caption{\label{fig:toth} Left panel (a) shows the ratio of the maxima of the
toroidal magnetic field strength and the equipartition value with
time;  (b)  shows the parity indexes evolution in the models.
We employ the running average to filter out the separate cycles.}
\end{figure}
To quantify the mirror symmetry type of the toroidal magnetic field
distribution relative to equator we introduce the parity index $P$:
\begin{eqnarray*}
P & = & \frac{E_{q}-E_{d}}{E_{q}+E_{d}},\\
E_{d} & = & \int\left(B\left(r_{0},\theta\right)-B\left(r_{0},\pi-\theta\right)\right)^{2}\sin\theta d\theta,\\
E_{q} & = & \int\left(B\left(r_{0},\theta\right)+B\left(r_{0},\pi-\theta\right)\right)^{2}\sin\theta d\theta,
\end{eqnarray*}
where $E_{d}$ and $E_{q}$ are the energies of the dipole-like and
quadruple-like modes, $r_{0}=0.9R_{\odot}$. We define
the dynamical quenching type 1 (QT1) to be govern by Eq.(\ref{eq:hel}),
and the dynamical quenching type 2 (QT2) - by Eq.~(\ref{eq:helcon2}).

\section{Results}
\begin{figure*}[t]
\noindent \begin{centering}
\includegraphics[width=0.7\textwidth]{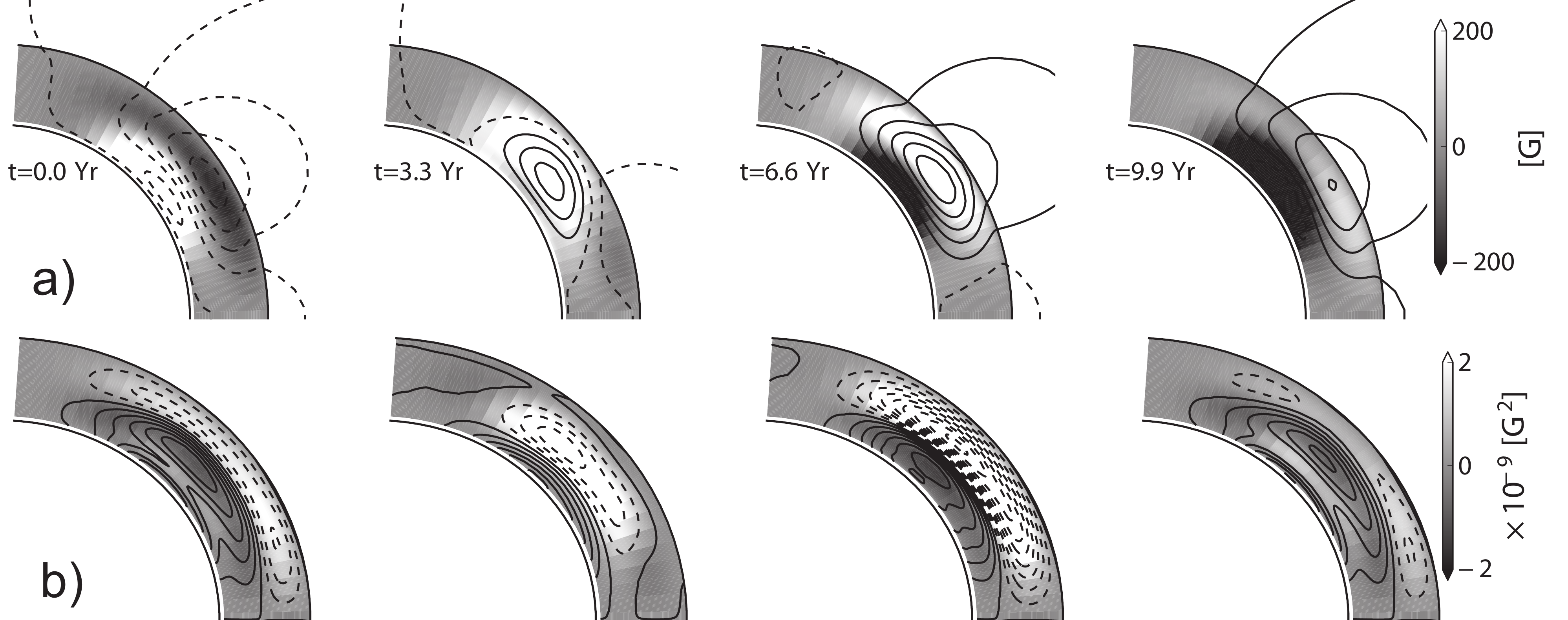} 
\par\end{centering}
\caption{ Snapshots of the magnetic field and helicity evolution inside the
North segment of the convection zone for the model QT2$_3$. Panel (a) shows the field
lines of the poloidal component of the mean magnetic field and the
toroidal magnetic field (varies $\pm700$G) by gray scale density
plot. The bottom panel, (b), shows the large-scale (density plot) and small-scale
magnetic helicity (contours) distributions. Both kinds of the magnetic
helicity vary in the same range of magnitude.\label{fig:Snapshots}}
\end{figure*}

Fig.~\ref{fig:toth} shows the long-term evolution of the maximum
of the large-scale magnetic field strength in the convection zone
and the parity in the models. The energy of the toroidal magnetic
fields in all models show the exponential grow in the beginning phase,
which has duration about 10 diffusive time of the system. The greater
initial magnetic field strength, the shorter duration of the exponential
grow phase. Two models QT1$_{1}$ and QT2$_{1}$ have rather small
diffusive fluxes of the helicity and the high magnetic Reynolds number
$R_{m}=10^{6}$. We consider them as the references. It is shown below,
see Figures \ref{fig:toth} and \ref{fig:helvar} that the model QT2$_{1}$
is not subjected to the catastrophic quenching while the model QT1$_{1}$
saturates the toroidal magnetic field strength to the level that is
much below the equipartition. The moderate diffusive flux with $\eta_{\chi}=0.01\eta_{T}$
(model QT1$_{3}$) does not alleviate the problem if the magnetic
helicity evolution is governed by the Eq.~(\ref{eq:hel}). In the
models QT1$_{2}$ and QT2$_{1,3}$ the saturation level of the toroidal
magnetic field strength is about $0.3B_{{\rm eq}}$, where $B_{{\rm eq}}$
is the equipartition level of the magnetic field strength. It is about
$0.5B_{eq}$ in the model QT2$_{2}$ which has $R_{m}=10^{3}$. The
saturation level in the QT2 types solar dynamo can be higher for the
the greater $C_{\alpha}$. This question needs a separate study. In
the model QT1$_{1}$ as well as in all the QT2 models the parity index
saturates to the dipole-like solution. In the QT1$_{2,3}$ models
the asymptotic stage is not clear and they need much longer runs.

The origin of difference in behaviour of the magnetic helicity evolution
in the models with QT1 and QT2 has been discussed recently by \citet{2012ApJ...748...51H}.
Taking into the dynamo equation (\ref{eq:dyn}), the corresponding
equation for the large-scale vector potential:
\begin{eqnarray}
\frac{\partial\overline{\mathbf{A}}}{\partial t} & = & \boldsymbol{\mathcal{E}}+\overline{\mathbf{U}}\times\overline{\mathbf{B}},\label{eq:A}
\end{eqnarray}
 where we assume the Coulomb gauge, we find the equation which governs
the large-scale helicity evolution: 
\begin{eqnarray}
\frac{\partial\left(\overline{\mathbf{A}}\cdot\overline{\mathbf{B}}\right)}{\partial
  t}\!\! & = &\!\!
2\boldsymbol{\mathcal{E}}\cdot\overline{\mathbf{B}}\!\!+\!\! 
\boldsymbol{\nabla}\cdot\left(\left(\boldsymbol{\mathcal{E}}\times\overline{\mathbf{A}}\right)
\!\!- \!\!
\mathbf{\overline{A}}\times\left(\overline{\mathbf{U}}\times\overline{\mathbf{B}}\right)\right).\label{AB}
\end{eqnarray}
Therefore, Eq. (\ref{eq:helcon2}) can be rewritten in form of Eq.(\ref{eq:int-cons}):
\begin{eqnarray}
\frac{\partial\overline{\chi}}{\partial t} & = &
-2\left(\boldsymbol{\mathcal{E}}\cdot\overline{\bm{B}}\right)-\frac{\overline{\chi}}{R_{m}\tau_{c}}+\boldsymbol{\nabla}\cdot\left(\eta_{\chi}\boldsymbol{\nabla}\bar{\chi}\right) \label{eq:hel-1}\\
&&-\eta\overline{\mathbf{B}}\cdot\mathbf{\overline{J}}
-\boldsymbol{\nabla}\cdot\left(\boldsymbol{\mathcal{E}}\times\overline{\mathbf{A}}\right).\nonumber
\end{eqnarray}
The term $\left(\boldsymbol{\mathcal{E}}\times\overline{\mathbf{A}}\right)$
consists of the counterparts of the sources magnetic helicity, which
are represented by $-2\boldsymbol{\mathcal{E}}\cdot\overline{\mathbf{B}}$,
and the fluxes which result from pumping of the large-scale magnetic
fields. The sources magnetic helicity in the term $-2\left(\boldsymbol{\mathcal{E}}\cdot\overline{\bm{B}}\right)$
are partly compensated in Eq(\ref{eq:hel-1}) by the counterparts
in $\left(\boldsymbol{\mathcal{E}}\times\overline{\mathbf{A}}\right)$.
This results in the spatially homogeneous quenching of the large-scale
magnetic generation and alleviation of the catastrophic quenching
problem. The last term in Eq(\ref{AB})
 % $\boldsymbol{\nabla}\cdot\left(\mathbf{\overline{A}}\times\left(\overline{\mathbf{U}}\times\overline{\mathbf{B}}\right)\right)$,
contains the transport of the large-scale magnetic helicity by the
large-scale flow.

Fig.~\ref{fig:Snapshots} shows the snapshots of the magnetic field
and magnetic helicity (large- and small-scale) evolution in the North
segment of the solar convection zone for the model QT2$_{3}$. We
observe the drift of the dynamo waves which are related to the large-scale
toroidal and poloidal fields towards the equator and towards the pole,
respectively. The distributions of the large- and small-scale magnetic
helicities show the correspondence in sign: positive to negative, and the
other way around, respectively. This is in agreement
with Eq.~(\ref{eq:helcon2}). It is seen that the negative sign of
the magnetic helicity follows the dynamo wave of the toroidal magnetic
field. This can be related to the so-called ``current helicity hemispheric
sign rule'' (negative/positive sign of helicity dominate in the North/South
hemisphere) which is suggested by the observations (see \citet{see1990SoPh,zetal10}
and references therein). The origin of the helicity sign rule has
been extensively studied in the dynamo theory
(e.g., \citealp{choud2004ApJ,2012ApJ...751...47Z}).
%\citep{choud2004ApJ,kps:06,soka06,pevts2007ASPC,pk11,2012ApJ...751...47Z}.
The similar snapshots for the QT1 model can be found in \citep{pk11}.
The main difference is that for the QT1 model all the changes in the
magnetic helicity evolution does not exactly follow to the dynamo
wave inside the convection zone as it is demonstrated for the QT2
model. 

\begin{figure*}
\noindent \begin{centering}
\includegraphics[width=0.8\textwidth]{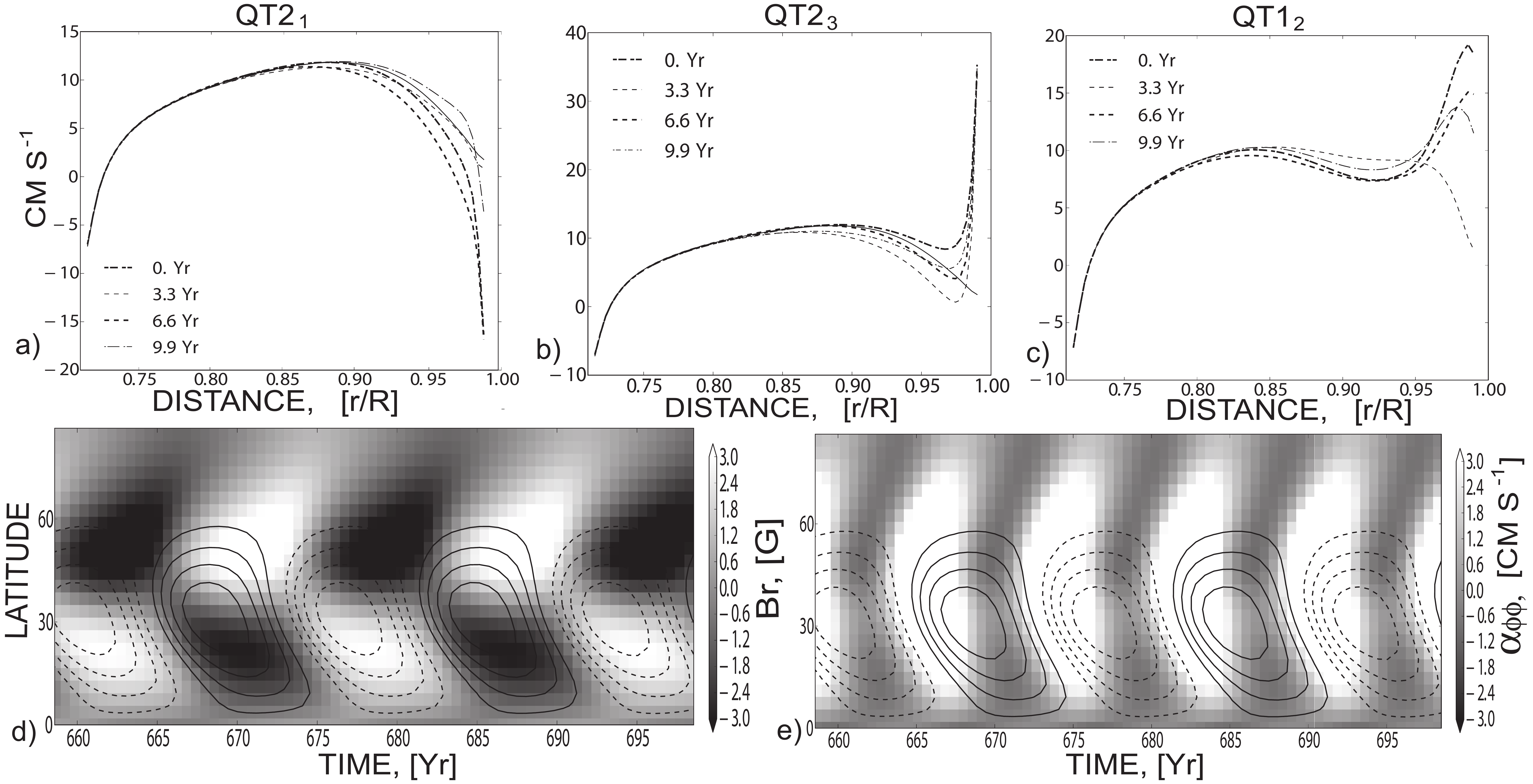}
\par\end{centering}
\caption{ First row, (a,b and c), shows variations of the $\alpha$ effect ($\alpha_{\phi\phi}$
component) profiles with the cycle at colatitude $\theta=45^{\circ}$,
for the model QT2$_{1}$, QT2$_{3}$ and QT1$_{2}$ (from left to
right). % Second row. (d,e and f) shows the same for the magnetic helicity.
The bottom,
panel (d) shows the time - latitude variations of the toroidal
field near the surface, $r=0.95R_{\odot}$,  (contours $\pm300$G)
and the radial magnetic field at the surface (density plot) for the model QT2$_{3}$;
panel (e) shows the same for the the toroidal field and the $\alpha$
effect ($\alpha_{\phi\phi}$ component)(density plot).\label{fig:helvar} }
\end{figure*}

Fig.~\ref{fig:helvar} shows variations of the radial profiles of
the $\alpha$ effect and magnetic helicity with the cycle and the
time-latitude diagrams for the dynamo model QT2$_{3}$. For all the
models the changes in the $\alpha$ effect are concentrated to the
top of the convection zone. This is due to the factor $\left(\overline{\rho}\ell^{2}\right)^{-1}$
in relation between the current and magnetic helicities. The difference
in the boundary at the top results in different distributions of the
$\alpha$ effect in the models QT2$_{1}$ and QT2$_{3}$. We would
like to notice that the QT1$_{2}$ and QT2$_{1}$ have the same boundary
conditions, however, the resulted distributions of the $\alpha$ effect
differ drastically. The time-latitude diagrams for the dynamo model
QT2$_{3}$ which are shown in Figure \ref{fig:helvar}, are in qualitative
agreement with observations. The same patterns are obtained in the
models QT2$_{1,2}$. We show the dynamical $\alpha$ effect as well.
The model shows that with the boundary conditions given by Eq.~(\ref{eq:bc2})
the $\alpha$ effect increases and has positive maxima at the growing
phase of the cycle and it decreases, having the negative minima at
the decaying phase of the cycle.
%  The models QT2$_{1,2}$ have similar
% time latitude diagrams. However, the dynamo wave of the toroidal magnetic
% field in the model QT2$_{3}$ penetrates about $2^{\circ}$ closer to
% equator than that in the models QT2$_{1,2}$ because the $\alpha$
% effect changes from positive to negative when the toroidal field get
% closer to the surface.

\section{Conclusion}

In the paper we studied the effect of the magnetic helicity conservation
in the mean-field solar dynamo model which is shaped by the subsurface
shear. The results show that the solar dynamo can operate in the wide
rage of the magnetic Reynolds number up to $10^{6}$ if conservation
of the total magnetic helicity is taken into account.

It was found that the boundary conditions for the magnetic helicity
influence the distribution of the $\alpha$ effect near the solar
surface. For example, the dynamo wave becomes closer to equator when
the diffusive flux of the total helicity is zero at the top of the
convection zone because the dynamical increase of the $\alpha$ effect
follows the dynamo wave (see Fig.~\ref{fig:Snapshots}, \ref{fig:helvar}).
The situation is opposite for the case when the diffusive flux is
dominated by the large-scale helicity. Thus, the parity breaking in
the solar dynamo can occur due to perturbations in the external boundary
for the magnetic helicity. However, these points remain an open field
for the future work.

\acknowledgements{}

V.P., D.S. and K.K. would like to acknowledge support from Visiting
Professorship Programme of Chinese Academy or Sciences 2009J2-12 and
thank NAOC of CAS for hospitality, as well as acknowledge
%  support
% from the RFBR under grants 12-02-00170-a, 10-02-00960-a, 10-02-00148-a,
the support of the Integration Project of SB RAS N 34, and support
of the state contracts 02.740.11.0576, 16.518.11.7065 of the Ministry
of Education and Science of Russian Federation. H.Z. would like to
acknowledge support from National Natural Science Foundation of China
grants: 41174153 and 10921303.

%\bibliographystyle{/home/vv/work/pap/aastex/8/new/apj}

% \bibliography{/home/vv/work/pap/dyn}

\end{document}